\documentclass[twocolumn,showpacs,preprintnumbers,amsmath,amssymb]{revtex4}

\usepackage{graphicx}
\usepackage{dcolumn}
\usepackage{bm}
\usepackage{html}

\newcommand{\poper}[1]{\frac{\partial}{\partial #1}}

\newcommand{\tplus}{{+}}
\newcommand{\tminus}{{-}}
\newcommand{\tpm}{{\pm}}

\newcommand{\Ts}{T}      
\newcommand{\Xs}{X}      

\newcommand{\e}{\epsilon} 

\newcommand{\yr}{y_r}
\newcommand{\yl}{y_l}

\newcommand{\fp}{f^\tplus}
\newcommand{\fm}{f^\tminus}

\newcommand{\vp}{v^\tplus}
\newcommand{\vm}{v^\tminus}

\newcommand{\yp}{y^\tplus}
\newcommand{\ym}{y^\tminus}

\newcommand{\mndx}{m} 
\newcommand{\ynm}{y_{n,\mndx}} 
\newcommand{\ynpom}{y_{n+1,\mndx}}
\newcommand{\ynpompo}{y_{n+1,\mndx+1}}
\newcommand{\ynmom}{y_{n-1,\mndx}}

\newcommand{\ynmmo}{y_{n,\mndx-1}}

\newcommand{\ynmpo}{y_{n,\mndx+1}}
\newcommand{\ynmompt}{y_{n-1,\mndx+2}}
\newcommand{\ynmompf}{y_{n-1,\mndx+4}}
\newcommand{\ynmpth}{y_{n,\mndx+3}}
\newcommand{\ynmpfi}{y_{n,\mndx+5}}

\newcommand{\ypnm}{\yp_{n,\mndx}}
\newcommand{\ypnmmo}{\yp_{n,\mndx-1}}
\newcommand{\ypnmpo}{\yp_{n,\mndx+1}}

\newcommand{\ypnmom}{\yp_{n-1,\mndx}}
\newcommand{\ypnmpth}{\yp_{n,\mndx+3}}
\newcommand{\ypnmpfi}{\yp_{n,\mndx+5}}

\newcommand{\ymnm}{\ym_{n,\mndx}}
\newcommand{\ymnmmo}{\ym_{n,\mndx-1}}
\newcommand{\ymnmpo}{\ym_{n,\mndx+1}}

\newcommand{\ymnmom}{\ym_{n-1,\mndx}}
\newcommand{\ymnmpth}{\ym_{n,\mndx+3}}
\newcommand{\ymnmpfi}{\ym_{n,\mndx+5}}

\newcommand{\yx}{y'}
\newcommand{\yxx}{y''}

\newcommand{\yt}{{\dot y}}
\newcommand{\ytt}{{\ddot y}}

\newcommand{\ypx}{y'^\tplus}
\newcommand{\ymx}{y'^\tminus}

\newcommand{\mytextwidth}{3in}

\newcommand{\captionwidth}{\mytextwidth}

\newcommand{\mycaptionparlabel}[2]{\parbox{\captionwidth}{\small\caption{\protect #1\label{\protect #2}}}}

\newcommand{\mystrut}{}
	
\newcommand{\myFigureToWidth}[3]{
	\begin{figure}[htbp]
	\centering
	\resizebox{#2}{!}{\includegraphics{eps/#1.eps}}
	\mycaptionparlabel{\protect #3}{fig:#1}
	\mystrut
	\end{figure}
}

{
\end{boxedminipage}
\parbox{\captionwidth}{{\small\caption{\currCap}\label{code:\currLabel}}}
\end{figure}
}

{
\hrule
\begin{minipage}{\captionwidth} 
{\small\caption{\currCap}\label{code:\currLabel}}
\end{minipage}
\end{boxedminipage}
\end{figure}
}

\usepackage{color} 

\newcommand{\myTexFigure}[2]{
\begin{figure}[htb]
\centering
\centerline{\input fig/#1.pstex_t}  
\mycaptionparlabel{\protect #2}{fig:#1}
\mystrut
\end{figure}
} 

\newcommand{\BEQN}{\begin{equation}}
\newcommand{\EEQN}{\end{equation}}
\newcommand{\BEQL}{\begin{equation}}
\newcommand{\EEQL}{\end{equation}}
\newcommand{\BEQA}{\begin{eqnarray}}
\newcommand{\EEQA}{\end{eqnarray}}
\newcommand{\BEAS}{\latexhtml{}{\par\begin{center}\begin{makeimage}}\begin{eqnarray*}} 
\newcommand{\EEAS}{\end{eqnarray*}\latexhtml{}{\end{makeimage}\end{center}\par}}
\newcommand{\beas}{\BEAS}
\newcommand{\eeas}{\EEAS}

\newcommand{\BA}{\begin{array}}
\newcommand{\EA}{\end{array}}

\newcommand{\beqn}{\begin{equation}}
\newcommand{\eeqn}{\end{equation}}
\newcommand{\beqa}{\begin{eqnarray}}
\newcommand{\eeqa}{\end{eqnarray}}

\newcommand{\ie}{\textit{i.e.}}

\newcommand{\eg}{\textit{e.g.}}

\providecommand{\isdeftext}{\buildrel {\scriptscriptstyle\Delta}\over{\scriptstyle=}}
\providecommand{\isdef}{\mathrel{\stackrel{\Delta}{=}}}

\newcommand{\eref}[1]{(\ref{eq:#1})}

\newcommand{\elabel}[1]{\protect\label{eq:#1}}

\providecommand{\fref}[1]{Fig.~\ref{fig:#1}}

\providecommand{\Fref}[1]{Figure~\ref{fig:#1}}

\providecommand{\seclabel}[1]{\label{sec:#1}}
\providecommand{\sref}[1]{\S\protect\ref{sec:#1}}

\newcommand{\kdef}[1]{\emph{#1}\index{#1|textbf}}  

\newcommand{\ksetcontext}[1]{\fixme{How to set K context?}}
\newcommand{\nnr}{\nonumber\\}

\newcommand{\xv}{\underline{x}}
\newcommand{\yv}{\underline{y}}
\newcommand{\uv}{\underline{u}}
\newcommand{\up}{\upsilon}
\newcommand{\upv}{\underline{\upsilon}}
\newcommand{\bv}{\underline{\beta}}
\newcommand{\g}{\gamma}
\newcommand{\gv}{\underline{\gamma}}
\newcommand{\gm}{\gamma^\tminus}
\newcommand{\gp}{\gamma^\tplus}
\newcommand{\gpm}{\gamma^\tpm}
\newcommand{\gvm}{\gv^\tminus}
\newcommand{\gvp}{\gv^\tplus}
\newcommand{\xK}{\xv_K}
\newcommand{\xW}{\xv_W}
\newcommand{\T}{\mathbf{T}}
\newcommand{\Ti}{\T^{-1}}
\newcommand{\A}{\mathbf{A}}
\newcommand{\B}{\mathbf{B}}
\newcommand{\C}{\mathbf{C}}

\newcommand{\AK}{\A_K}
\newcommand{\AKt}{{\tilde \A}_K}

\newcommand{\nacc}[2]{\mbox{$\stackrel{{\scriptscriptstyle #1}}{#2}$}}
\newcommand{\Att}{\nacc{\vdash}{\A}}
\newcommand{\AKtt}{\Att_K}
\newcommand{\AWtt}{\Att_W}
\newcommand{\AKttt}{\nacc{\vdash\!\!\dashv}{\A}_K}
\newcommand{\AWttt}{\nacc{\vdash\!\!\dashv}{\A}_W}
\newcommand{\AKtttt}{\nacc{\rightleftharpoons}{\A}_K}

\newcommand{\BK}{\B_K}
\newcommand{\CK}{\C_K}

\newcommand{\AW}{\A_W}
\newcommand{\AWt}{{\tilde \A}_W}
\newcommand{\BW}{{\B_W}}
\newcommand{\CW}{\C_W}

\newcommand{\zi}{z^{-1}}
\newcommand{\zmt}{z^{-2}}
\newcommand{\Oscr}{\mathcal{O}}

\newcommand{\ints}{{\bf Z}}

\newcommand{\db}{\bm{\delta}}
\newcommand{\Db}{\mathbf{\Delta}}

\newcommand{\ejoT}{e^{j\omega\Ts}}
\newcommand{\abs}[1]{\left|#1\right|}

\newcommand{\BNUM}{\begin{enumerate}}
\newcommand{\ENUM}{\end{enumerate}}
\newcommand{\BIT}{\begin{itemize}}
\newcommand{\EIT}{\end{itemize}}

\newcommand{\Appl}{Appl.}

\newcommand{\Int}{Int.}
\newcommand{\Journal}{J.}
\newcommand{\JournalOfThe}{J.}
\newcommand{\JournalP}{J.}

\newcommand{\Soc}{Soc.}
\newcommand{\Proc}{Proc.}
\newcommand{\Conf}{Conf.}
\newcommand{\ConfOn}{Conf.}

\newcommand{\Symp}{Symp.}
\newcommand{\Dept}{Dept.}
\newcommand{\Ama}{Amer.}
\newcommand{\SocOf}{Soc.\ of}
\newcommand{\Acoust}{Acoust.}

\begin{document}


\title{On the Equivalence of the Digital Waveguide and \\
Finite Difference Time Domain Schemes}

\author{Julius O. Smith III}
\homepage{http://ccrma.stanford.edu/~jos/}
\email{jos@ccrma.stanford.edu}
\affiliation{%
Center for Computer Research in Music and Acoustics (CCRMA)\\
Stanford University\\
Stanford, CA 94305
}%

\date{\today}

\begin{abstract}
It is known that the digital waveguide (DW) method for solving the
wave equation numerically on a grid can be manipulated into the form
of the standard finite-difference time-domain (FDTD) method (also
known as the ``leapfrog'' recursion).  This paper derives a simple
rule for going in the other direction, that is, converting the state
variables of the FDTD recursion to corresponding wave variables in a
DW simulation.  Since boundary conditions and initial values are more
intuitively transparent in the DW formulation, the simple means of
converting back and forth can be useful in initializing and
constructing boundaries for FDTD simulations.
\end{abstract}



\maketitle

\section{Introduction} 
\seclabel{intro}

The digital waveguide (DW) method has been used for many years to
provide highly efficient algorithms for musical sound synthesis based
on physical models \cite{MADW,SmithDWMMI}.  For a much longer time,
finite-difference time-domain (FDTD) schemes have been used to
simulate more general situations at generally higher cost
\cite{Ruiz,ChaigneFDA,ChaigneAndAskenfeltBothParts,BensaEtAlJASA03}.
In recent years, there has been interest in relating these methods to
each other \cite{ErkutAndMattiISMA02} and in combining them for more
general simulations.  For example, modular hybrid methods have been
devised which interconnect DW and FDTD simulations by means of a
\emph{KW-pipe} \cite{MattiMohonk03,MattiAndErkut04}.  The basic idea
of the KW-pipe adaptor is to convert the ``Kirchoff variables'' of the
FDTD, such as string displacement, velocity, etc., to ``wave
variables'' of the DW.  The W variables are regarded as the
traveling-wave components of the K variables.

In this paper, we present an alternative to the KW pipe.  Instead of
converting K variables to W variables, or vice versa, in the time
domain, conversion formulas are derived with respect to the current
state as a function of spatial coordinates.  As a result, it becomes
simple to convert any instantaneous state configuration from FDTD to
DW form, or vice versa.  Thus, instead of providing the necessary
time-domain filter to implement a KW pipe converting traveling-wave
components to physical displacement of a vibrating string, say, one
may alternatively set the displacement variables instantaneously to
the values corresponding to a given set of traveling-wave components
in the string model.  Another benefit of the formulation is an exact
physical interpretation of arbitrary initial conditions and
excitations in the K-variable FDTD method.  Since the DW formulation
is exact in principle (though bandlimited), while the FDTD is
approximate, even in principle, it can be argued that the true
physical interpretation of the FDTD method is that given by the DW
method.  Since both methods generate the same evolution of state from
a common starting point, they may only differ in computational
expense, numerical sensitivity, and in the details of supplying
initial conditions and boundary conditions.

\section{Ideal String Wave Equation} 
\seclabel{dwtofd}
For definiteness, let's consider simulating the ideal vibrating
string, as shown in \fref{Fphysicalstring}.

\myFigureToWidth{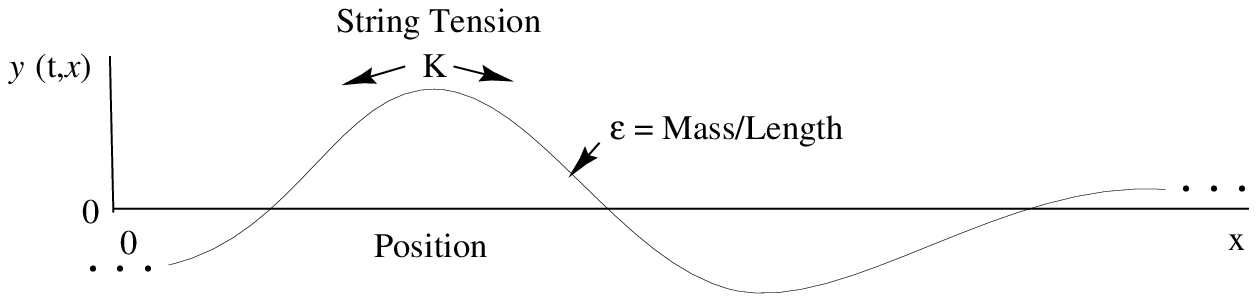}{\mytextwidth}{The ideal vibrating string.}

The \kdef{wave equation} for the ideal (lossless, linear, flexible) 
vibrating string depicted in \fref{Fphysicalstring} is given by
\beqn
K\yxx = \e\ytt
\elabel{Ewe}
\eeqn
where
\beqn
\begin{array}{rclrcl}
  K & \isdef & \mbox{string tension} & \qquad y & \isdef & y(t,x) \nonumber \\
  \e & \isdef & \mbox{linear mass density} 
& \yt & \isdef & \poper{t}y(t,x) \nonumber \\
   y & \isdef & \mbox{string displacement} & \yx & \isdef & \poper{x}y(t,x) \nonumber
\end{array}
\eeqn 
and ``$\isdef$'' means ``is defined as.''  The wave equation is
derived, \eg, in \cite{Morse}.

\subsection{Finite Difference Time Domain Scheme}

Using centered finite difference approximations (FDA) for the
second-order partial derivatives, we obtain a \emph{finite difference
scheme} for the ideal wave equation
\cite{Strikwerda,Moin01}:
\beqa
\ytt(t,x) &\approx& \frac{y(t+\Ts,x) - 2 y(t,x) + y(t-\Ts,x) }{\Ts^2}\\
\yxx(t,x) &\approx& \frac{y(t,x+\Xs) - 2 y(t,x) + y(t,x-\Xs) }{\Xs^2}
\elabel{fda}
\eeqa
where $\Ts$ is the time sampling interval, and $\Xs$ is a spatial
sampling interval. 

Substituting the FDA into the wave equation, choosing $\Xs=c\Ts$,
where $c\isdeftext \sqrt{K/\e}$ is sound speed (normalized to $c=1$
below), and sampling at times $t=n\Ts$ and $x=m\Xs$, we obtain
the following explicit finite difference scheme for the string
displacement: 
\beqn 
y(n+1,m) = y(n,m+1) + y(n,m-1) - y(n-1,m)
\label{eq:EFDTD} 
\eeqn 
where the sampling intervals $\Ts$ and $\Xs$ have been normalized to
1.  To initialize the recursion at time $n=0$, past values are needed
for all $m$ (all points along the string) at time instants $n=-1$ and
$n=-2$.  Then the string position may be computed for all $m$ by
\eref{EFDTD} for $n=0,1,2,\ldots\,$.  This has been called the
\emph{FDTD} or leapfrog finite difference scheme \cite{Essl04}.

\subsection{Digital Waveguide Scheme}

We now derive the digital waveguide formulation by \emph{sampling} the
\emph{traveling-wave} solution to the wave equation.  It is easily
checked that the lossless 1D wave equation $K\yxx =\e\ytt$ is solved
by any string shape $y$ which travels to the left or right with speed
$c \isdeftext \sqrt{K/\e}$ \cite{dAlembert}.  Denote
\emph{right-going} traveling waves in general by $\yr(t-x/c)$ and
\emph{left-going} traveling waves by $\yl(t+x/c)$, where $\yr$ and
$\yl$ are assumed twice-differentiable.  Then, as is well known, the
general class of solutions to the lossless, one-dimensional,
second-order wave equation can be expressed as
\beqn
y(t,x) = \yr\left(t-\frac{x}{c}\right) + \yl\left(t+\frac{x}{c}\right).
\elabel{Etwd}
\eeqn
Sampling these traveling-wave solutions yields
\beqa
  y(n\Ts,m\Xs) &=& \yr(n\Ts - m\Xs/c) + \yl(n\Ts + m\Xs/c) \nnr
               &=& \yr[(n-m)\Ts] + \yl[(n+m)\Ts] \nnr
               &\isdef& \yp(n-m) + \ym(n+m)
\elabel{sampledda}
\eeqa
where a ``$+$'' superscript denotes a ``right-going'' traveling-wave
component, and ``$-$'' denotes propagation to the ``left''.  This
notation is similar to that used for acoustic-tube modeling of speech
\cite{MG}.


\myFigureToWidth{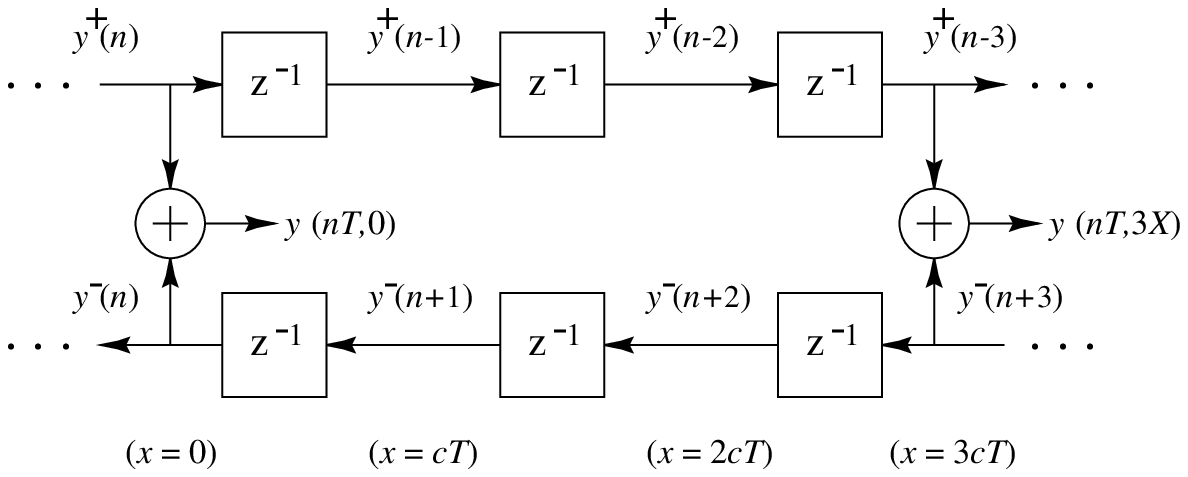}{\mytextwidth}{Digital simulation of the ideal, lossless waveguide
with observation points at $x=0$ and $x=3\Xs=3c\Ts$.  (The symbol
``$z^{-1}$'' denotes a one-sample delay.)}

\Fref{fideal} shows a signal flow diagram for the computational model
of \eref{sampledda}, which is often called a digital waveguide model
(for the ideal string in this case) \cite{SmithPMUDW,SmithDWMMI}.
Note that, by the sampling theorem, it is an exact model so long as
the initial conditions and any ongoing additive excitations are
bandlimited to less than half the temporal sampling rate $f_s =
1/\Ts$ \cite[Appendix G]{MDFT}.

Note also that the position along the string, $x_m = m\Xs = m c\Ts$
meters, is laid out from left to right in the diagram, giving a
physical interpretation to the horizontal direction in the diagram,
even though spatial samples have been eliminated from explicit
consideration. (The arguments of $\yp$ and $\ym$ have physical units
of time.)

The left- and right-going traveling wave components are summed to
produce a physical output according to 
\beqn 
y(n\Ts,m\Xs) = \yp(n-m) + \ym(n+m) 
\elabel{Esumout} 
\eeqn 
In \fref{fideal}, ``transverse displacement outputs'' have been
arbitrarily placed at $x=0$ and $x=3\Xs$.  The diagram is similar to
that of well known ladder and lattice digital filter structures
\cite{MG}, except for the delays along the upper rail, the absence of
scattering junctions, and the direct physical interpretation.

\subsection{FDTD and DW Equivalence}

The FDTD and DW recursions both compute time updates by forming fixed
linear combinations of past state.  As a result, each can be described
in ``state-space form'' \cite[Appendix E]{JOSFP} by a constant matrix
operator, the ``state transition matrix'', which multiplies the state
vector at the current time to produce the state vector for the next
time step.  The FDTD operator propagates K variables while the DW
operator propagates W variables.  We may show equivalence by (1)
defining a one-to-one transformation which will convert K variables to
W variables or vice versa, and (2) showing that given any common
initial state for both schemes, the state transition matrices compute
the same next state in both cases.

The next section shows that the linear transformation from W to K
variables, 
\beqn
y(n,m) = \yp(n-m) + \ym(n+m),
\elabel{map}
\eeqn
for all $n$ and $m$, sets up a one-to-one linear transformation
between the K and W variables.  Assuming this holds, it only
remains to be shown that the DW and FDTD schemes preserve mapping
\eref{map} after a state transition from one time to the next.  While
this has been shown previously \cite{SmithMohonkBook98}, we repeat
the derivation here for completeness.  We also provide a state-space
analysis reaching the same conclusion in \sref{ssf}.

From \fref{fideal}, it is clear that the DW scheme preserves mapping
\eref{map} by definition.  For the FDTD scheme, we expand the
right-hand of \eref{EFDTD} using \eref{map} and verify that the
left-hand side also satisfies the map, \ie, that $y(n+1,m) =
\yp(n+1-m) + \ym(n+1+m)$ holds:
\beas
y(n+1,m) &=& y(n,m+1) + y(n,m-1) - y(n-1,m) \\
        &=&       \yp(n-m-1) + \ym(n+m+1) \\
        && + \yp(n-m+1) + \ym(n+m-1) \\
        && - \yp(n-m-1) - \ym(n+m-1) \\
        &=&  \ym(n+m+1) + \yp(n-m+1) \\
        &=&  \yp[(n+1)-m] + \ym[(n+1)+m] \\
        &\isdef& y(n+1,m) \nonumber
\eeas
%
%
Since the DW method propagates sampled (bandlimited) solutions to the
ideal wave equation without error, it follows that the FDTD method
does the same thing, despite the relatively crude approximations made
in \eref{fda}.  In particular, it is known that the FDA introduces
artificial damping when applied to first order partial derivatives
arising in lumped, mass-spring systems \cite{SmithDWMMI}.

The equivalence of the DW and FDTD state transitions extends readily
to the DW mesh \cite{VanDuyneAndSmithMesh,SmithDWMMI} which is
essentially a lattice-work of DWs for simulating membranes and
volumes.  The equivalence is more important in higher dimensions
because the FDTD formulation requires less computations per node than
the DW approach in higher dimensions (see \cite{BeesonEtAlDAFX04} for
some quantitative comparisons).

Even in one dimension, the DW and finite-difference methods have
unique advantages in particular situations \cite{MattiMohonk03}, and
as a result they are often combined together to form a hybrid
traveling-wave/physical-variable simulation
\cite{PitteroffAndWoodhouse98b,PitteroffAndWoodhouse98c,Matti02,ErkutAndMatti02,ErkutAndMattiISMA02,MattiSMAC03,ArvindhAndSmithMohonk03,BeesonEtAlDAFX04}.

\section{State Transformations} 
\seclabel{fdtodw}

In previous work, time-domain adaptors (digital filters) converting
between K variables and W variables have been devised
\cite{MattiMohonk03}.  In this section, an alternative approach is
proposed.  Mapping \eref{map} gives us an immediate conversion from W
to K state variables, so all we need now is the inverse map for any
time $n$.  This is complicated by the fact that non-local spatial
dependencies can go indefinitely in one direction along the string, as
we will see below.  We will proceed by first writing down the
conversion from W to K variables in matrix form, which is easy to do,
and then invert that matrix.  For simplicity, we will consider the
case of an infinitely long string.

To initialize a K variable simulation for starting at time $n+1$, we
need initial spatial samples at all positions $m$ for two successive
times $n-1$ and $n$.  From this state specification, the FDTD scheme
\eref{EFDTD} can compute $y(n+1,m)$ for all $m$, and so on for
increasing $n$.  In the DW model, all state variables are defined as
belonging to the same time $n$, as shown in \fref{wglossless}.

\vfil
\penalty0

\myTexFigure{wglossless}{DW flow diagram.}

From \eref{Esumout}, and referring to the notation defined in
\fref{wglossless}, we may write the conversion from W to K variables
as
\beqa
\ynmpo &=& \ypnmpo + \ymnmpo\nnr
\ynmmo &=& \ypnmmo + \ymnmmo\nnr
\ynmom &=& \ypnmom + \ymnmom\nnr
       &=& \ypnmpo + \ymnmmo
\elabel{ynmom}
\eeqa
where the last equality follows from the traveling-wave behavior
(see \fref{wglossless}).  

\myTexFigure{stencil}{Stencil of the FDTD scheme.}

\Fref{stencil} shows the so-called ``stencil'' of the FDTD scheme.
The larger circles indicate the state at time $n$ which can be used to
compute the state at time $n+1$.  The filled and unfilled circles
indicate membership in one of two interleaved grids \cite{BilbaoT}. To
see why there are two interleaved grids, note that when $m$ is even,
the update for $\ynpom$ depends only on odd $m$ from time $n$ and even
$m$ from time $n-1$.  Since the two W components of $\ynmom$ are converted to
two W components at time $n$ in \eref{ynmom}, we have that the update for
$\ynpom$ depends only on W components from time $n$ and positions
$m\pm1$.
Moving to the next position update, for $\ynpompo$, the state used is
independent of that used for $\ynpom$, and the W components used are
from positions $m$ and $m+2$.  As a result of these observations, we
see that we may write the state-variable transformation separately for
even and odd $m$, \eg,
\begin{widetext}
\beqn
\left[\!
\begin{array}{c}
\vdots \\
\ynmmo \\
\ynmom \\
\ynmpo \\
\ynmompt \\
\ynmpth\\
\ynmompf \\
\ynmpfi \\
\vdots \\
\end{array}
\!\right]
\!=\!
\left[\!
\begin{array}{cccccccccc}
\ddots &  & \vdots & \vdots & \vdots & \vdots & \vdots & \vdots & \vdots & 0 \\
\cdots & 1 & 1 & 0 & 0 & 0 & 0 & 0 & 0 & \cdots\\
\cdots & 0 & 1 & 1 & 0 & 0 & 0 & 0 & 0 & \cdots\\
\cdots & 0 & 0 & 1 & 1 & 0 & 0 & 0 & 0 & \cdots\\
\cdots & 0 & 0 & 0 & 1 & 1 & 0 & 0 & 0 & \cdots\\
\cdots & 0 & 0 & 0 & 0 & 1 & 1 & 0 & 0 & \cdots\\
\cdots & 0 & 0 & 0 & 0 & 0 & 1 & 1 & 0 & \cdots\\
\cdots & 0 & 0 & 0 & 0 & 0 & 0 & 1 & 1 & \cdots\\
  0   & \vdots & \vdots & \vdots & \vdots & \vdots & \vdots & \vdots  & \ddots 
\end{array}
\!\right]
\left[\!
\begin{array}{c}
\vdots \\
\ypnmmo \\
\ymnmmo \\
\ypnmpo \\
\ymnmpo \\
\ypnmpth \\
\ymnmpth \\
\ypnmpfi \\
\ymnmpfi \\
\vdots 
\end{array}
\!\right].
\elabel{wtok}
\eeqn
\end{widetext}
Denote the linear transformation operator by $\T$ and the K and W state vectors
by $\xK$ and $\xW$, respectively.  Then \eref{wtok} can be restated as
\beqn
\xK = \T \xW.
\elabel{cc}
\eeqn
The operator $\T$ can be recognized as the Toeplitz operator
associated with the linear, shift-invariant filter $H(z)=1+\zi$.
While the present context is not a simple convolution since $\xW$ is
not a simple time series, the inverse of $\T$ corresponds to the
Toeplitz operator associated with
\[
H(z) = \frac{1}{1+\zi} = 1 - \zi + z^{-2} - z^{-3} + \cdots.
\]
Therefore, we may easily write down the inverted transformation:
\begin{widetext}
\beqn
\left[\!
\begin{array}{c}
\vdots \\
\ypnmmo \\
\ymnmmo \\
\ypnmpo \\
\ymnmpo \\
\ypnmpth \\
\ymnmpth \\
\ypnmpfi \\
\ymnmpfi \\
\vdots 
\end{array}
\!\right]
\!=\!
\left[\!
\begin{array}{crrrrrrrrc}
\ddots &  & \vdots & \vdots & \vdots & \vdots 
          & \vdots & \vdots & \vdots & \pm1 \\
\cdots & 1 & -1 &  1 & -1 &  1 & -1 &  1 & -1 & \cdots\\
\cdots & 0 &  1 & -1 &  1 & -1 &  1 & -1 &  1 & \cdots\\
\cdots & 0 &  0 &  1 & -1 &  1 & -1 &  1 & -1 & \cdots\\
\cdots & 0 &  0 &  0 &  1 & -1 &  1 & -1 &  1 & \cdots\\
\cdots & 0 &  0 &  0 &  0 &  1 & -1 &  1 & -1 & \cdots\\
\cdots & 0 &  0 &  0 &  0 &  0 &  1 & -1 &  1 & \cdots\\
\cdots & 0 &  0 &  0 &  0 &  0 &  0 &  1 & -1 & \cdots\\
  0   & \vdots & \vdots & \vdots & \vdots & \vdots & \vdots & \vdots 
  & \ddots 
\end{array}
\!\right]
\left[\!
\begin{array}{c}
\vdots \\
\ynmmo \\
\ynmom \\
\ynmpo \\
\ynmompt \\
\ynmpth\\
\ynmompf \\
\ynmpfi \\
\vdots \\
\end{array}
\!\right]
\elabel{ktow}
\eeqn
\end{widetext}
The case of the finite string is identical to that of the infinite
string when the matrix $\T$ is simply ``cropped'' to a finite square size
(leaving an isolated 1 in the lower right corner); in such cases,
$\Ti$ as given above is simply cropped to the same size, retaining its
upper triangular $\pm1$ structure.  Another interesting set of cases
is obtained by inserting a 1 in the lower-left corner of the cropped
$\T$ matrix to make it \emph{circulant}; in these cases, the $M\times
M$ matrix $\Ti$ contains $\pm1/2$ in every position for even $M$, and
is singular for odd $M$ (when there is one zero eigenvalue).


\section{Examples} 
\seclabel{examples}
\subsection{Localized Displacement Excitations}
Whenever two adjacent components of $\xK$ are initialized with equal
amplitude, only a single $W$-variable will be affected.  For example,
the initial conditions (for time $n+1$)
\beas
y_{n,m-1} &=& 1\\
y_{n-1,m} &=& 1 
\eeas
will initialize only $\ymnmmo$, a solitary left-going pulse 
of amplitude 1 at time
$n=0$, as can be seen from \eref{ktow} by adding the leftmost columns
explicitly written for $\Ti$.  Similarly, the initialization
\beas
y_{n-1,m-2} &=& 1\\
y_{n,m-1} &=& 1
\eeas
gives rise to an isolated right-going pulse $\ypnmmo$, corresponding
to the leftmost column of $\Ti$ plus the first column on the left not
explicitly written in \eref{ktow}.  The superposition of these two
examples corresponds to a physical impulsive excitation at time 0 and
position $m-1$:
\beqa
y_{n-1,m-2} &=& 1\nnr
y_{n,m-1} &=& 2\nnr
y_{n-1,m} &=& 1
\elabel{dispimp}
\eeqa
Thus, the impulse starts out with amplitude 2 at time 0 and position
$m-1$, and afterwards, impulses of amplitude 1 propagate away to the
left and right along the string.  

In summary, we see that to excite a single sample of displacement
traveling in a single-direction, we must excite equally a pair of
adjacent colums in $\Ti$.  This corresponds to equally weighted
excitation of K-variable pairs the form $(y_{n,m},y_{n-1,m\pm1})$.

Note that these examples involved only one of the two interleaved
computational grids. Shifting over an odd number of spatial samples to
the left or right would involve the other grid, as would shifting time
forward or backward an odd number of samples.

\subsection{Localized Velocity Excitations}

Initial velocity excitations are straightforward in the DW paradigm,
but can be less intuitive in the FDTD domain.  It is well known that
velocity in a displacement-wave DW simulation is determined by the
\emph{difference} of the right- and left-going waves
\cite{SmithPMUDW}.  Specifically, initial velocity waves $v^{\pm}$ can
be computed from from initial displacement waves $y^\pm$ by spatially
differentiating $y^\pm$ to obtain traveling \emph{slope waves}
$\yx^\pm$, multiplying by minus the tension $K$ to obtain \emph{force
waves}, and finally dividing by the wave impedance $R=\sqrt{K\e}$ to
obtain velocity waves:
\beqa
\vp &=& -c\ypx = \frac{\fp}{R}\nnr
\vm &=& \;c\ymx = -\frac{\fm}{R},
\elabel{veldef}
\eeqa
where $c=\sqrt{K/\e}$ denotes sound speed.  The initial string
velocity at each point is then $v(n\Ts,m\Xs)=\vp(n-m)+\vm(n+m)$.  (A
more direct derivation can be based on differentiating \eref{Etwd}
with respect to $x$ and solving for velocity traveling-wave
components, considering left- and right-going cases separately at
first, and arguing the general case by superposition.)

We can see from \eref{ktow} that such asymmetry can be caused by
unequal weighting of $y_{n,m}$ and $y_{n,m\pm1}$.  For example, the
initialization
\beas
y_{n-1,m+1} &=& +1\\
y_{n-1,m} &=& -1
\eeas
corresponds to an impulse \emph{velocity} excitation at position
$m+1/2$.  In this case, both interleaved grids are excited.

\subsection{More General Velocity Excitations}
\seclabel{mgve}

From \eref{ktow}, it is clear that initializing any single K variable
$\ynm$ corresponds to the initialization of an infinite number of W
variables $\ypnm$ and $\ymnm$.  That is, a single K variable $\ynm$
corresponds to only a single column of $\Ti$ for only one of the
interleaved grids.  For example, 
referring to \eref{ktow},
initializing the K variable
$\ynmom$ to -1 at time $n$ (with all other $\ynm$ intialized to 0)
corresponds to the W-variable initialization
\beas
\yp_{n,m-(2\mu+1)}&=&+1, \quad \mu =0,1,2,\cdots\\
\ym_{n,m-(2\mu+1)}&=&-1, \quad \mu =0,1,2,\cdots
\eeas
with all other W variables being initialized to zero.  
In view of earlier remarks, this corresponds to an impulsive velocity
excitation on only one of the two subgrids.  A schematic
depiction from $\mu = m-5$ to $m+5$ of the W variables at time $n$ is as
follows:
\beqn
\begin{array}{crrrrr|rrrrrrc}
\cdots &  1 &  0 &  1 &  0 &  1 & 0 & 0 & 0 & 0 & 0 & 0 & \cdots\\
\cdots & -1 &  0 & -1 &  0 & -1 & 0 & 0 & 0 & 0 & 0 & 0 & \cdots\\
\hline\\[-1em]
\cdots &  0 &  0 &  0 &  0 &  0 & 0 & 0 & 0 & 0 & 0 & 0 & \cdots\\
      &    &    &    &    &    & m &   &   &   &   & \mu & \rightarrow
\end{array}
\eeqn
Below the solid line is the sum of the left- and right-going
traveling-wave components, \ie, the corresponding K variables at time
$n$.  The vertical lines divide positions $\mu=m-1$ and $\mu=m$.
The initial displacement is zero everywhere at time $n$, 
consistent with an initial velocity excitation.  
At times $\nu=n+1,n+2,n+3,n+4$, the configuration evolves as follows:
\beqn
\begin{array}{crrrrr|rrrrrrc}
\cdots &  0 &  1 &  0 &  1 & 0 & 1 & 0 & 0 & 0 & 0 & 0 & \cdots\\
\cdots &  0 & -1 &  0 & -1 & 0 & 0 & 0 & 0 & 0 & 0 & 0 & \cdots\\
\hline\\[-1em]
\cdots &  0 &  0 &  0 &  0 & 0 & 1 & 0 & 0 & 0 & 0 & 0 & \cdots
\end{array}
\eeqn
\beqn
\begin{array}{crrrrr|rrrrrrc}
\cdots &  1 &  0 &  1 & 0 & 1 & 0 & 1 & 0 & 0 & 0 & 0 & \cdots\\
\cdots & -1 &  0 & -1 & 0 & 0 & 0 & 0 & 0 & 0 & 0 & 0 & \cdots\\
\hline\\[-1em]
\cdots &  0 &  0 &  0 & 0 & 1 & 0 & 1 & 0 & 0 & 0 & 0 & \cdots
\end{array}
\eeqn
\beqn
\begin{array}{crrrrr|rrrrrrc}
\cdots & 0 &  1 & 0 & 1 & 0 & 1 & 0 & 1 & 0 & 0 & 0 & \cdots\\
\cdots & 0 & -1 & 0 & 0 & 0 & 0 & 0 & 0 & 0 & 0 & 0 & \cdots\\
\hline\\[-1em]
\cdots & 0 &  0 & 0 & 1 & 0 & 1 & 0 & 1 & 0 & 0 & 0 & \cdots
\end{array}
\eeqn
\beqn
\begin{array}{crrrrr|rrrrrrc}
\cdots &  1 & 0 & 1 & 0 & 1 & 0 & 1 & 0 & 1 & 0 & 0 & \cdots\\
\cdots & -1 & 0 & 0 & 0 & 0 & 0 & 0 & 0 & 0 & 0 & 0 & \cdots\\
\hline\\[-1em]
\cdots &  0 & 0 & 1 & 0 & 1 & 0 & 1 & 0 & 1 & 0 & 0 & \cdots
\end{array}
\eeqn
The sequence $[\dots,1,0,1,0,1,\dots]$ consists of a dc
(zero-frequency) component with amplitude $1/2$, plus a sampled
sinusoid of amplitude $1/2$ oscillating at half the sampling rate
$f_s=1/\Ts$.  The dc component is physically correct for an initial
velocity point-excitation (a spreading square pulse on the string is
expected).  However, the component at $f_s/2$ is usually regarded as
an artifact of the finite difference scheme.  From the DW
interpretation of the FDTD scheme, which is an exact, bandlimited
physical interpretation, we see that physically the component at
$f_s/2$ comes about from initializing velocity on only one of the two
interleaved subgrids of the FDTD scheme.  Any asymmetry in the
excitation of the two interleaved grids will result in excitation of
this frequency component.

Due to the independent interleaved subgrids in the FDTD algorithm, it
is nearly always non-physical to excite only one of them, as the above
example makes clear.  It is analogous to illuminating only every other
pixel in a digital image.  However, joint excitation of both grids may be
accomplished either by exciting adjacent spatial samples at the same
time, or the same spatial sample at successive times instants.

In addition to the W components being non-local, they can demand a
larger dynamic range than the K variables.  For example, if the entire
semi-infinite string for $m<0$ is initialized with velocity $2\Xs/\Ts$,
the initial displacement traveling-wave components look as follows:
\beqn
\begin{array}{crrrrrr|rrrrrc}
\cdots &  6 &  5 &  4 &  3 &  2 &  1 & 0 & 0 & 0 & 0 & 0 & \cdots\\
\cdots & -6 & -5 & -4 & -3 & -2 & -1 & 0 & 0 & 0 & 0 & 0 & \cdots\\
\hline\\[-1em]
\cdots &  0 &  0 &  0 &  0 &  0 & 0 & 0 & 0 & 0 & 0 & 0 & \cdots
\end{array}
\eeqn
and the variables evolve forward in time as follows:
\beqn
\begin{array}{crrrrrr|rrrrrc}
\cdots &  7 &  6 &  5 &  4 &  3 & 2 & 1 & 0 & 0 & 0 & 0 & \cdots\\
\cdots & -5 & -4 & -3 & -2 & -1 & 0 & 0 & 0 & 0 & 0 & 0 & \cdots\\
\hline\\[-1em]
\cdots &  2 &  2 &  2 &  2 &  2 & 2 & 1 & 0 & 0 & 0 & 0 & \cdots
\end{array}
\eeqn
\beqn
\begin{array}{crrrrrr|rrrrrc}
\cdots &  8 &  7 &  6 &  5 & 4 & 3 & 2 & 1 & 0 & 0 & 0 & \cdots\\
\cdots & -4 & -3 & -2 & -1 & 0 & 0 & 0 & 0 & 0 & 0 & 0 & \cdots\\
\hline\\[-1em]
\cdots &  4 &  4 &  4 &  4 & 4 & 3 & 2 & 1 & 0 & 0 & 0 & \cdots
\end{array}
\eeqn
\beqn
\begin{array}{crrrrrr|rrrrrc}
\cdots &  9 &  8 &  7 & 6 & 5 & 4 & 3 & 2 & 1 & 0 & 0 & \cdots\\
\cdots & -3 & -2 & -1 & 0 & 0 & 0 & 0 & 0 & 0 & 0 & 0 & \cdots\\
\hline\\[-1em]
\cdots &  6 &  6 &  6 & 6 & 5 & 4 & 3 & 2 & 1 & 0 & 0 & \cdots
\end{array}
\eeqn
Thus, the left semi-infinite string moves upward at a constant
velocity of 2, while a ramp spreads out to the left and right of
position $\mu=m$ at speed $c$, as expected physically.  By
\eref{wtok}, the corresponding initial FDTD state for this case is
\beas
y_{n,\mu} &=&0, \quad  \mu\in\ints\\
y_{n-1,m-1} &=& -1,\\
y_{n-1,\mu} &=& -2, \quad \mu<m-1,
\eeas
where $\ints$ denotes the set of all integers.
While the FDTD excitation is also not local, of course, it is
bounded for all $\mu$.

Since the traveling-wave components of initial velocity excitations
are generally non-local in a displacement-based simulation, as
illustrated in the preceding examples, it is often preferable to use
velocity waves (or force waves) in the first place \cite{SmithDWMMI}.

Another reason to prefer force or velocity waves is that displacement
inputs are inherently impulsive. To see why this is so, consider that
any physically correct driving input must effectively exert some
finite force on the string, and this force is free to change
arbitrarily over time. The ``equivalent circuit'' of the infinitely
long string at the driving point is a ``dashpot'' having real,
positive resistance $2R=2\sqrt{K\e}$.  The applied force $f(t)$ can be
divided by $2R$ to obtain the velocity $v(t)$ of the string driving
point, and this velocity is free to vary arbitrarily over time,
proportional to the applied force.  However, this velocity must be
\emph{time-integrated} to obtain a displacement $y(t)$.  Therefore,
there can be \emph{no instantaneous displacement response to a finite
driving force}.  In other words, any instantaneous effect of an input
driving signal on an output displacement sample is non-physical except
in the case of a massless system.  Infinite force is required to move
the string instantaneously.  In sampled displacement simulations, we
must interpret displacement changes as resulting from time-integration
over a sampling period.  As the sampling rate increases, any
physically meaningful displacement driving signal must converge to
zero.

\subsection{Additive Inputs}

Instead of initial conditions, ongoing input signals can be defined
analogously.  For example, feeding an input signal $u_n$ into the FDTD
via
\beqa
y_{n,m-1} &=& y_{n,m-1}  + u_{n-1}\nnr
y_{n,m} &=& y_{n,m}  + 2u_n \nnr
y_{n,m+1} &=& y_{n,m+1}  + u_{n-1}
\elabel{dispinp}
\eeqa
corresponds to physically driving a single sample of string
displacement at position $m$.  This is the spatially distributed
alternative to the temporally distributed solution of feeding an input
to a single displacement sample via the filter $H(z)=1-z^{-2}$ as
discussed in \cite{MattiMohonk03}.  


\subsection{Physical Interpretation of $H(z)=1-z^{-2}$}

As shown above, driving a single displacement sample $\ynm$ in the
FDTD corresponds to driving a velocity input at position $m$ on two
alternating subgrids over time.  Therefore, the filter $H(z)=1-z^{-2}$
acts as the filter $H(z)=1-z^{-1}$ on either subgrid alone---a
first-order difference.  Since displacement is being simulated, velocity
inputs must be numerically integrated.  The first-order difference can
be seen as canceling this integration, thereby converting a velocity
input to a displacement input, as in \eref{dispinp}.  

\section{State Space Formulation}
\seclabel{ssf}

In this section, we will summarize and extend the above discussion by
means of a \emph{state space analysis} \cite{Kailath80}.

\subsection{FDTD State Space Model}

Let $\xK(n)$ denote the FDTD state for one of the two subgrids at time
$n$, as defined by \eref{cc}.  The other subgrid is handled
identically and will not be considered explicitly.  In fact, the other
subgrid can be dropped altogether to obtain a \emph{half-rate,
staggered grid} scheme \cite{BilbaoT,Fornberg98}.  However, boundary
conditions and input signals will couple the subgrids, in general.  To
land on the same subgrid after a state update, it is necessary to
advance time by two samples instead of one.  The state-space model for
one subgrid of the FDTD model of the ideal string may then be written
as
\beqa
\xK(n+2) &=& \AK\, \xK(n) + \BK\, \uv(n+2)\nnr
\yv(n) &=& \CK\, \xK(n).
\elabel{ssmk}
\eeqa
To avoid the issue of boundary conditions for now, we will continue
working with the infinitely long string.  As a result, the state
vector $\xK(n)$ denotes a point in a space of countably infinite
dimensionality.  A proper treatment of this case would be in terms of
operator theory \cite{NaylorAndSell82}.  However, matrix notation is
also clear and will be used below.  Boundary conditions are taken up
in \sref{bc}.

When there is a general input signal vector $\uv(n)$, it is necessary to
augment the input matrix $\BK$ to accomodate contributions over both
time steps.  This is because inputs to positions $m\pm1$ at time $n+1$
affect position $m$ at time $n+2$.  Henceforth, we assume $\BK$ and
$\uv$ have been augmented in this way.  Thus, if there are $q$ input
signals $\upv(n)\isdeftext[\up_i(n)]$, $i=1,\ldots,q$, driving the full
string state through weights $\bv_m\isdeftext[\beta_{m,i}]$,
$m\in\ints$, the vector $\uv(n)=$ is of dimension $2q\times 1$:
\[
\uv(n+2) = 
\left[\!
\begin{array}{c}
\upv(n+2)\\
\upv(n+1)
\end{array}
\!\right]
\]
When there is only one physical input, as is typically assumed
for plucked, struck, and bowed strings, then $q=1$ and $\uv$ is
$2\times1$.  The matrix $\BK$ weights these inputs before they are
added to the state vector for time $n+2$, and its entries are derived
in terms of the $\beta_{m,i}$ coefficients below.

$\CK$ forms the output signal as an arbitrary linear combination of
states.  To obtain the usual displacement output for the subgrid,
$\CK$ is the matrix formed from the identity matrix by deleting every
other row, thereby retaining all displacement samples at time $n$ and
discarding all displacement samples at time $n-1$ in the state vector
$\xK(n)$:
\[
\underbrace{\left[\!
\begin{array}{c}
\vdots \\
y_{n,m-2} \\
y_{n,m} \\
y_{n,m+2} \\
y_{n,m+4}\\
\vdots 
\end{array}
\!\right]}_{\yv(n)}
\!=\!
\underbrace{\left[\!
\begin{array}{ccccccccccc}
 & \vdots & \vdots & \vdots & \vdots & \vdots & \vdots & \vdots  & \vdots\\
\cdots & 1 & 0 & 0 & 0 & 0 & 0 & 0 & 0 & \cdots\\
\cdots & 0 & 0 & 1 & 0 & 0 & 0 & 0 & 0 & \cdots\\
\cdots & 0 & 0 & 0 & 0 & 1 & 0 & 0 & 0 & \cdots\\
\cdots & 0 & 0 & 0 & 0 & 0 & 0 & 1 & 0 & \cdots\\
 & \vdots & \vdots & \vdots & \vdots & \vdots & \vdots & \vdots  & \vdots
\end{array}
\!\right]}_{\CK}
\underbrace{\left[\!
\begin{array}{c}
\vdots \\
y_{n,m-2} \\
y_{n-1,m-1} \\
y_{n,m} \\
y_{n-1,m+1} \\
y_{n,m+2} \\
y_{n-1,m+3} \\
y_{n,m+4}\\
\vdots 
\end{array}
\!\right]}_{\xK(n)}
\]
The state transition matrix $\AK$ may be obtained by first performing
a one-step time update,
\[
y_{n+2,m} = y_{n+1,m-1}+y_{n+1,m+1}-y_{n,m}+\bv_m^T \upv(n+2), 
\]
and then expanding the two $n+1$ terms in terms of the state at time $n$:
\beqa
y_{n+1,m-1} &=& y_{n,m-2}+y_{n,m}-y_{n-1,m-1}+\bv_{m-1}^T\upv(n+1)
		\nonumber\\[5pt]
y_{n+1,m+1} &=& y_{n,m}+y_{n,m+2}-y_{n-1,m+1}+\bv_{m+1}^T\upv(n+1)\nnr
&& \elabel{oddm}
\eeqa
The intra-grid state update for even $m$ is then given by
\beqa
\lefteqn{y_{n+2,m}} \nnr
&=& y_{n,m-2} - y_{n-1,m-1} + y_{n,m} - y_{n-1,m+1} + y_{n,m+2}\nnr
&& \quad +\; \bv_m^T \upv(n+2) + (\bv_{m-1}+\bv_{m+1})^T\upv(n+1)\nnr
&=&
\begin{array}{c}
\left[1, -1, 1, -1, 1\right]\\
\vspace{0.5in}
\end{array}
\left[\!
\begin{array}{l}
y_{n,m-2}\\
y_{n-1,m-1}\\
y_{n,m}\\
y_{n-1,m+1}\\
y_{n,m+2}\\
y_{n-1,m+3}\\
\end{array}
\!\right]
 	  \nnr
& & + \left[\bv_m^T \quad (\bv_{m-1}+\bv_{m+1})^T \right]
\left[\!
\begin{array}{l}
\upv(n+2)\\
\upv(n+1)
\end{array}
\!\right].
\elabel{igsu}
\eeqa
For odd $m$, the update in \eref{oddm} is used.  Thus, every other row
of $\AK$, for time $n+2$, consists of the vector $[1,-1,1,-1,1]$
preceded and followed by zeros.  Successive rows for time $n+2$ are
shifted right two places.  The rows for time $n+1$ consist of the
vector $[1,-1,1]$ aligned similarly:
{\small
\[
\underbrace{\left[\!
\begin{array}{l}
\qquad\vdots\\
y_{n+1,m-1}\\
y_{n+2,m}\\
y_{n+1,m+1}\\
y_{n+2,m+2}\\
\qquad\vdots
\end{array}
\!\right]}_{\xK(n+2)}
\!\leftarrow\!
\underbrace{\left[\!
\begin{array}{rrrrrrrrrrr}
       & \vdots & \vdots & \vdots & \vdots & \vdots & \vdots & \vdots \\
\cdots &  1 & -1 & 1 &  0 &  0 &  0 &  0 & \cdots\\
\cdots &  1 & -1 & 1 & -1 &  1 &  0 &  0 & \cdots\\
\cdots &  0 &  0 & 1 & -1 &  1 &  0 &  0 & \cdots\\
\cdots &  0 &  0 & 1 & -1 &  1 & -1 &  1 & \cdots\\
       &  \vdots & \vdots & \vdots & \vdots & \vdots & \vdots & \vdots 
\end{array}
\!\right]}_{\AK}
\!
\underbrace{\left[\!
\begin{array}{l}
\qquad\vdots\\
y_{n,m-2}\\
y_{n-1,m-1}\\
y_{n,m}\\
y_{n-1,m+1}\\
y_{n,m+2}\\
y_{n-1,m+3}\\
\qquad\vdots
\end{array}
\!\right]}_{\xK(n)}
\]
}
From \eref{igsu} we also see that the input matrix $\BK$ is given 
as defined in the following expression:
\[
\underbrace{\left[\!
\begin{array}{l}
\qquad\vdots\\
y_{n+1,m-1}\\
y_{n+2,m}\\
y_{n+1,m+1}\\
y_{n+2,m+2}\\
y_{n+1,m+3}\\
y_{n+2,m+4}\\
y_{n+1,m+5}\\
\qquad\vdots
\end{array}
\!\right]}_{\xK(n+2)}
\leftarrow
\underbrace{\left[\!
\begin{array}{cc}
\vdots & \vdots \\
0 & \bv_{m-1}^T      		\\[5pt]
\bv_m^T & \quad \bv_{m-1}^T + \bv_{m+1}^T \\[5pt]
0 & \bv_{m+1}^T     		\\[5pt]
\bv_{m+2}^T & \bv_{m+1}^T + \bv_{m+3}^T   \\[5pt]
0 & \bv_{m+3}^T       		\\[5pt]
\bv_{m+4}^T & \bv_{m+3}^T + \bv_{m+5}^T \\[5pt]
0 & \bv_{m+5}^T       		\\[5pt]
\vdots & \vdots
\end{array}
\!\right]}_{\BK}
\underbrace{\left[\!
\begin{array}{c}
\upv(n+2)\\
\upv(n+1)
\end{array}
\!\right]}_{\uv(n+2)}.
\]

\subsection{DW State Space Model}

As discussed in \sref{fdtodw}, the traveling-wave decomposition
\eref{Etwd} defines a linear transformation \eref{cc} from the DW
state to the FDTD state:
\beqn
\xK = \T\, \xW
\elabel{xform}
\eeqn
Since $\T$ is invertible, it qualifies as a linear transformation
for performing a \emph{change of coordinates} for the state space.
Substituting \eref{xform} into the FDTD 
state space model \eref{ssmk} gives
\beqa
\T\,\xW(n+2) &=& \AK\, \T\,\xW(n) + \BK\, \uv(n+2)\elabel{ssmws}\\
\yv(n) &=& \CK\, \T\,\xW(n).
\elabel{ssmwo}
\eeqa
Multiplying through \eref{ssmws} by $\Ti$ gives a new state-space
representation of the same dynamic system
which we will show is in fact the DW model of \fref{wglossless}:
\beqa
\xW(n+2) &=& \AW\, \xW(n) + \BW\, \uv(n+2)\nnr
\yv(n) &=& \CW\, \xW(n)
\eeqa
where
\beqa
\AW &\isdef& \Ti\AK\,\T\nnr
\BW &\isdef& \Ti\BK\nnr
\CW &\isdef& \CK\,\T
\elabel{ssmdw}
\eeqa
To verify that the DW model derived in this manner is the
computation diagrammed in \fref{wglossless}, we may write down the
state transition matrix for one subgrid from the figure to obtain
the permutation matrix $\AW$,
{\small
\beqn
\underbrace{\left[\!
\begin{array}{l}
\qquad\vdots \\
\yp_{n+2,m-2} \\
\ym_{n+2,m-2} \\
\yp_{n+2,m} \\
\ym_{n+2,m} \\
\yp_{n+2,m+2} \\
\ym_{n+2,m+2} \\
\qquad\vdots
\end{array}
\!\right]}_{\xW(n+2)}
\!\leftarrow\!
\underbrace{\left[\!
\begin{array}{cccccccccccc}
\cdots & \vdots & \vdots & \vdots & \vdots & \vdots & \vdots & \vdots & \vdots & \vdots & \vdots \\
\cdots &  1 & 0 & 0 & 0 & 0 & 0 & 0 & 0 & 0 & 0 & \cdots\\
\cdots &  0 & 0 & 0 & 0 & 0 & 1 & 0 & 0 & 0 & 0 & \cdots\\
\cdots &  0 & 0 & 1 & 0 & 0 & 0 & 0 & 0 & 0 & 0 & \cdots\\
\cdots &  0 & 0 & 0 & 0 & 0 & 0 & 0 & 1 & 0 & 0 & \cdots\\
\cdots &  0 & 0 & 0 & 0 & 1 & 0 & 0 & 0 & 0 & 0 & \cdots\\
\cdots &  0 & 0 & 0 & 0 & 0 & 0 & 0 & 0 & 0 & 1 & \cdots\\
\cdots & \vdots & \vdots & \vdots & \vdots & \vdots & \vdots & \vdots & \vdots & \vdots & \vdots 
\end{array}
\!\right]}_{\AW}
\!
\underbrace{\left[\!
\begin{array}{l}
\quad\vdots \\
\yp_{n,m-4} \\
\ym_{n,m-4} \\
\yp_{n,m-2} \\
\ym_{n,m-2} \\
\yp_{n,m} \\
\ym_{n,m} \\
\yp_{n,m+2} \\
\ym_{n,m+2} \\
\yp_{n,m+4} \\
\ym_{n,m+4} \\
\quad\vdots
\end{array}
\!\right]}_{\xW(n)}
\elabel{AWinf}
\eeqn
}
and displacement output matrix $\CW$:
\[
\underbrace{\left[\!
\begin{array}{c}
\vdots \\
y_{n,m-2} \\
y_{n,m} \\
y_{n,m+2} \\
y_{n,m+4}\\
\vdots 
\end{array}
\!\right]}_{\yv(n)}
=
\underbrace{\left[\!
\begin{array}{cccccccccc}
 & \vdots & \vdots & \vdots & \vdots & \vdots & \vdots & \vdots & \vdots & \\
\cdots & 1 & 1 & 0 & 0 & 0 & 0 & 0 & 0 & \cdots\\
\cdots & 0 & 0 & 1 & 1 & 0 & 0 & 0 & 0 & \cdots\\
\cdots & 0 & 0 & 0 & 0 & 1 & 1 & 0 & 0 & \cdots\\
\cdots & 0 & 0 & 0 & 0 & 0 & 0 & 1 & 1 & \cdots\\
 & \vdots & \vdots & \vdots & \vdots & \vdots & \vdots & \vdots & \vdots &
\end{array}
\!\right]}_{\CW}
\underbrace{\left[\!
\begin{array}{l}
\quad\vdots \\
\yp_{n,m-2} \\
\ym_{n,m-2} \\
\yp_{n,m} \\
\ym_{n,m} \\
\yp_{n,m+2} \\
\ym_{n,m+2} \\
\yp_{n,m+4} \\
\ym_{n,m+4} \\
\quad\vdots
\end{array}
\!\right]}_{\xW(n)}
\]


\subsubsection{DW Displacement Inputs}

We define general DW inputs as follows:
\beqa
\ypnm &=& \yp_{n-1,m-1} + (\gvp_m)^T \upv(n)\\
\ymnm &=& \ym_{n-1,m+1} + (\gvm_m)^T \upv(n)
\eeqa
The $m$th $2q\times2$ block of the input matrix $\BW$ driving state 
components $[\yp_{n+2,m},\ym_{n+2,m}]^T$ and multiplying
$[\upv(n+2)^T,\upv(n+1)^T]^T$ is then given by
\beqn
\left(\BW\right)_m =
\left[\!
\begin{array}{cc}
(\gvp_m)^T & (\gvp_{m-1})^T \\[5pt]
(\gvm_m)^T & (\gvm_{m+1})^T
\end{array}
\!\right].
\elabel{gammas}
\eeqn
Typically, input signals are injected equally to the left and right
along the string, in which case 
\[
\gvp_m = \gvm_m \isdef \gv_m.
\]
Physically, this corresponds to applied forces at a single,
non-moving, string position over time.  The state update with this
simplification appears as
\[
\underbrace{\left[\!
\begin{array}{c}
\vdots\\
\yp_{n+2,m}\\[5pt]
\ym_{n+2,m}\\
\vdots
\end{array}
\!\right]}_{\xW(n+2)}
=\AW\xW(n) + 
\underbrace{\left[\!
\begin{array}{cc}
\vdots & \vdots\\
\gv_m^T & \gv_{m-1}^T \\[5pt]
\gv_m^T & \gv_{m+1}^T \\[5pt]
\vdots & \vdots
\end{array}
\!\right]}_{\BW}
\underbrace{\left[\!
\begin{array}{c}
\upv(n+2)\\
\upv(n+1)
\end{array}
\!\right]}_{\uv(n+2)}.
\]
Note that if there are no inputs driving the adjacent subgrid
($\gv_{m-1}=\gv_{m+1}=0$), such as in a half-rate staggered grid
scheme, the input reduces to
\[
\xW(n+2) = \AW\xW(n) + 
\underbrace{\left[\!
\begin{array}{c}
\vdots\\
\gv_{m-2}^T \\[5pt]
\gv_{m-2}^T \\[5pt]
\gv_m^T \\[5pt]
\gv_m^T \\[5pt]
\gv_{m+2}^T \\[5pt]
\gv_{m+2}^T \\[5pt]
\vdots
\end{array}
\!\right]}_{\BW}
\upv(n+2).
\]

To show that the directly obtained FDTD and DW state-space models
correspond to the same dynamic system, it remains to verify that
$\AW=\Ti\AK\,\T$.  It is somewhat easier to show that
\beas
\T\,\AW &=& \AK\,\T\\
&=&
\left[\!
\begin{array}{cccccccccccc}
& \vdots & \vdots & \vdots & \vdots & \vdots & \vdots & \vdots & \vdots & \vdots & \vdots\\
\cdots & 1 & 0 & 0 & 0 &  0 & 1 &  0 & 0 & 0 & 0 & \cdots\\
\cdots & 0 & 0 & 1 & 0 &  0 & 1 &  0 & 0 & 0 & 0 & \cdots\\
\cdots & 0 & 0 & 1 & 0 &  0 & 0 &  0 & 1 & 0 & 0 & \cdots\\
\cdots & 0 & 0 & 0 & 0 &  1 & 0 &  0 & 1 & 0 & 0 & \cdots\\
& \vdots & \vdots & \vdots & \vdots & \vdots & \vdots & \vdots & \vdots & \vdots & \vdots 
\end{array}
\!\right].  
\eeas
A straightforward calculation verifies that the above identity holds,
as expected.  One can similarly verify $\CW=\CK\,\T$, as expected.
The relation $\BW=\Ti\,\BK$ provides a recipe for translating any
choice of input signals for the FDTD model to equivalent inputs for
the DW model, or vice versa.
For example, in the scalar input case ($q=1$), the DW input-weights
$\BW$ become FDTD input-weights $\BK$ according to
\[
\left[\!
\begin{array}{l}
\qquad\vdots\\
y_{n+1,m-1}\\
y_{n+2,m}\\
y_{n+1,m+1}\\
y_{n+2,m+2}\\
\qquad\vdots
\end{array}
\!\right]
\!
\leftarrow
\!
\underbrace{\left[\!
\begin{array}{cc}
\vdots & \vdots\\
\gp_m +\gm_{m-1} \,&\, \gp_{m-1}+\gm_{m-1} \\[5pt]
\gp_m +\gm_m     \,&\, \gp_{m-1}+\gm_{m+1} \\[5pt]
\gm_m +\gp_{m+1} \,&\, \gp_{m+1}\gm_{m+1} \\[5pt]
\gp_{m+2}+\gm_{m+2} \,&\, \gp_{m+1}+\gm_{m+3} \\[5pt]
\vdots & \vdots
\end{array}
\!\right]}_{\BK}
\!
\left[\!
\begin{array}{c}
\upv(n+2)\\
\upv(n+1)
\end{array}
\!\right]
\]
The left- and right-going input-weight superscripts indicate the
origin of each coefficient.  Setting $\gp_m=\gm_m$ results in
\beqn
\BK = 
\left[\!
\begin{array}{cc}
\vdots & \vdots\\
\g_m +\g_{m-1} \,&\, 2\g_{m-1}         \\[5pt]
2\g_m          \,&\, \g_{m-1}+\g_{m+1} \\[5pt]
\g_m +\g_{m+1} \,&\, 2\g_{m+1}        \\[5pt]
2\g_{m+2}      \,&\, \g_{m+1}+\g_{m+3} \\[5pt]
\vdots & \vdots
\end{array}
\!\right]
\elabel{gdif}
\eeqn
Finally, when $\g_m=1$ and $\g_{\mu}=0$ for all $\mu\neq m$, we obtain the
result familiar from \eref{dispinp}:
\[
\BK=
\left[\!
\begin{array}{cc}
\vdots & \vdots\\
0 & 1 \\
2 & 0 \\
0 & 1 \\
\vdots & \vdots
\end{array}
\!\right]
\]
Similarly, setting $\gpm_{\mu}=0$ for all $\mu\neq m+1$, the weighting
pattern $(1,2,1)$ appears in the second column, shifted down one row.
Thus, $\BK$ in general (for physically stationary displacement inputs)
can be seen as the superposition of weight patterns $(1,2,1)$ in the
left column for even $m$, and the right column for odd $m$ (the other
subgrid), where the $2$ is aligned with the driven sample.  
This is the general collection of displacement inputs.

\subsubsection{DW Non-Displacement Inputs}
\seclabel{dwddi}

Since a displacement input at position $m$ corresponds to
symmetrically exciting the right- and left-going traveling-wave
components $\yp_m$ and $\ym_m$, it is of interest to understand what
it means to excite these components \emph{antisymmetrically}.  As
discussed in \sref{mgve}, an antisymmetric excitation of
traveling-wave components can be interpreted as a \emph{velocity}
excitation. It was noted that localized velocity excitations in the
FDTD generally correspond to non-localized velocity excitations in the
DW, and that velocity in the DW is proportional to the \emph{spatial
derivative} of the difference between the left-going and right-going
traveling displacement-wave components (see \eref{veldef}).  More
generally, the antisymmetric component of displacement-wave excitation
can be expressed in terms of any wave variable which is linearly
independent relative to displacement, such as acceleration, slope,
force, momentum, and so on.  Since the state space of a vibrating
string (and other mechanical systems) is traditionally taken to be
position and velocity, it is perhaps most natural to relate the
antisymmetric excitation component to velocity.

In practice, the simplest way to handle a velocity input $v_m(n)$ in a
DW simulation is to first pass it through a first-order integrator of the
form
\beqn
H(z) = \frac{1}{1-\zi} = 1 + \zi + z^{-2} + \cdots
\elabel{integ}
\eeqn
to convert it to a displacement input.  By the equivalence of the DW
and FDTD models, this works equally well for the FDTD model.  However,
in view of \sref{mgve}, this approach does not take full advantage of
the ability of the FDTD scheme to provide localized velocity inputs
for applications such as simulating a piano hammer strike.  The FDTD
provides such velocity inputs for ``free'' while the DW requires the
external integrator \eref{integ}.

Note, by the way, that these ``integrals'' (both that done internally
by the FDTD and that done by \eref{integ}) are merely sums over
discrete time---not true integrals. As a result, they are exact only
at dc (and also trivially at $f_s/2$, where the output amplitude is
zero).  Discrete sums can also be considered exact integrators for
impulse-train inputs---a point of view sometimes useful when
interpreting simulation results.  For normal bandlimited signals,
discrete sums most accurately approximate integrals in a neighborhood
of dc.  The KW-pipe filter $H(z)=1-\zmt$ has analogous properties.



\subsubsection{Input Locality}

The DW state-space model is given in terms of the FDTD state-space
model by \eref{ssmdw}.  The similarity transformation matrix $\T$ is
bidiagonal, so that $\CK$ and $\CW=\CK\,\T$ are both approximately
diagonal when the output is string displacement for all $m$.  However,
since $\Ti$ given in \eref{ktow} is upper triangular, the input matrix
$\BW=\Ti\BK$ can replace sparse input matrices $\BK$ with only
half-sparse $\BW$, unless successive columns of $\Ti$ are equally
weighted, as discussed in \sref{examples}.  We can say that local
K-variable excitations may correspond to \emph{non-local} W-variable
excitations.  From \eref{gammas} and \eref{gdif}, we see that
\emph{displacement inputs are always local in both systems}.
Therefore, local FDTD and non-local DW excitations can only occur when
a variable dual to displacement is being excited, such as velocity.
If the external integrator \eref{integ} is used, all inputs are
ultimately displacement inputs, and the distinction disappears.

\subsection{Boundary Conditions}
\seclabel{bc}

The relations of the previous section do not hold exactly when the
string length is finite.  A finite-length string forces consideration
of \emph{boundary conditions}.  In this section, we will introduce
boundary conditions as perturbations of the state transition matrix.
In addition, we will use the DW-FDTD equivalence to obtain physically
well behaved boundary conditions for the FDTD method.


Consider an ideal vibrating string with $M=8$
spatial samples.  This is a sufficiently large number to make clear
most of the repeating patterns in the general case.  Introducing
boundary conditions is most straightforward in the DW paradigm.  We
therefore begin with the order 8 DW model, for which the state vector
(for the $0$th subgrid) will be
\[
\xW(n) = 
\left[\!
\begin{array}{l}
\yp_{n,0}\\
\ym_{n,0}\\
\yp_{n,2}\\
\ym_{n,2}\\
\yp_{n,4}\\
\ym_{n,4}\\
\yp_{n,6}\\
\ym_{n,6}\\
\end{array}
\!\right].
\]
The displacement output matrix is given by
\[
\CW = 
\left[\!
\begin{array}{ccccccccccc}
 1 & 1 & 0 & 0 & 0 & 0 & 0 & 0 \\
 0 & 0 & 1 & 1 & 0 & 0 & 0 & 0 \\
 0 & 0 & 0 & 0 & 1 & 1 & 0 & 0 \\
 0 & 0 & 0 & 0 & 0 & 0 & 1 & 1 
\end{array}
\!\right]
\]
and the input matrix $\BW$ is an arbitrary $M\times 2q$ matrix.  We
will choose a scalar input signal $u(n)$ driving the displacement
of the second spatial sample with unit gain:
\[
\BW
=
\left[\!
\begin{array}{cc}
 0  &  0 \\
 0  &  0 \\
1/2 & 1/2 \\
1/2 & 1/2 \\
 0  &  0 \\
 0  &  0 \\
 0  &  0 \\
 0  &  0 
\end{array}
\!\right]
\]
The state transition matrix $\AW$ is obtained by reducing \eref{AWinf}
to finite order in some way, thereby introducing boundary conditions.

\subsubsection{Resistive Terminations}

Let's begin with simple ``resistive'' terminations at the string
endpoints, resulting in the reflection coefficient $g$ at each end of
the string, where $|g| \leq 1$ corresponds to nonnegative (passive)
termination resistances \cite{SmithDWMMI}.  Inspection of
\fref{wglossless} makes it clear that terminating the left endpoint may be
accomplished by setting
\[
\yp_{n,0} = g_l\ym_{n,0},
\]
and the right termination corresponds to
\[
\ym_{n,6} = g_r\yp_{n,6}.
\]
By allowing an additional two samples of round-trip delay in each
endpoint reflectance (one sample in the chosen subgrid), we can
implement these reflections within the state-transition matrix:
\beqn
\AWt = 
\left[\!
\begin{array}{ccccccccccc}
 0 & g_l & 0 & 0 &  0 & 0 & 0 & 0 \\
 0 & 0 & 0 & 1 &  0 & 0 & 0 & 0 \\
 1 & 0 & 0 & 0 &  0 & 0 & 0 & 0 \\
 0 & 0 & 0 & 0 &  0 & 1 & 0 & 0 \\
 0 & 0 & 1 & 0 &  0 & 0 & 0 & 0 \\
 0 & 0 & 0 & 0 &  0 & 0 & 0 & 1 \\
 0 & 0 & 0 & 0 &  1 & 0 & 0 & 0 \\
 0 & 0 & 0 & 0 &  0 & 0 & g_r & 0
\end{array}
\!\right]
\eeqn
The simplest choice of state transformation matrix $\T$ is obtained
by cropping it to size $M\times M$:
\[
\T\isdef 
\left[\!
\begin{array}{ccccccccccc}
 1 & 1 & 0 & 0 &  0 & 0 & 0 & 0 \\
 0 & 1 & 1 & 0 &  0 & 0 & 0 & 0 \\
 0 & 0 & 1 & 1 &  0 & 0 & 0 & 0 \\
 0 & 0 & 0 & 1 &  1 & 0 & 0 & 0 \\
 0 & 0 & 0 & 0 &  1 & 1 & 0 & 0 \\
 0 & 0 & 0 & 0 &  0 & 1 & 1 & 0 \\
 0 & 0 & 0 & 0 &  0 & 0 & 1 & 1 \\
 0 & 0 & 0 & 0 &  0 & 0 & 0 & 1
\end{array}
\!\right]
\]
An advantage of this choice is that its inverse $\Ti$ is similarly
a simple cropping of the $M=\infty$ case.  However, the corresponding
FDTD system is not so elegant:
\beas
\AKt &\isdef& \T\AWt\Ti\\
& = &
\left[\!
\begin{array}{ccccccccccc}
 0 &  g_l & -g_l & h_l & -h_l  & h_l & -h_l & h_l \\
 1 & -1 & 1 &  0 & 0 &  0 &  0 &  0 \\
 1 & -1 & 1 & -1 & 1 &  0 &  0 &  0 \\
 0 &  0 & 1 & -1 & 1 &  0 &  0 &  0 \\
 0 &  0 & 1 & -1 & 1 & -1 &  1 &  0 \\
 0 &  0 & 0 &  0 & 1 & -1 &  1 &  0 \\
 0 &  0 & 0 &  0 & 1 & -1 &  h_r & -h_r \\
 0 &  0 & 0 &  0 & 0 &  0 &  g_r &  -g_r 
\end{array}
\!\right],
\eeas
where $h_l\isdef 1+g_l$ and $h_r\isdef 1+g_r$.  We see that the left
FDTD termination is \emph{non-local} for $g\neq -1$, while the right
termination is local (to two adjacent spatial samples) for all $g$.
This can be viewed as a consequence of having ordered the FDTD state
variables as $[y_{n,m},y_{n-1,m+1},\ldots]$ instead of
$[y_{n-1,m},y_{n,m+1},\ldots]$.  Choosing the other ordering
interchanges the endpoint behavior.  Call these orderings Type I and
Type II, respectively. Then $\T_{II}=\T_I^T$; that is, the similarity
transformation matrix $\T$ is transposed when converting from Type I
to Type II or vice versa.  By anechoically coupling a Type I FDTD
simulation on the right with a Type II simulation on the left,
general resistive terminations may be obtained on both ends which are
localized to two spatial samples.



In nearly all musical sound synthesis applications, at least one of
the string endpoints is modeled as rigidly clamped at the ``nut''.
Therefore, since the FDTD, as defined here, most naturally provides
a clamped endpoint on the left, with more general localized terminations
possible on the right, we will proceed with this case for simplicity in what
follows. Thus, we set $g_l=-1$ and obtain
\beas
\AKtt &\isdef&
\left[\!
\begin{array}{ccccccccccc}
 0 & -1 & 1 & 0 &  0 &  0 &  0 &  0 \\
 1 & -1 & 1 &  0 & 0 &  0 &  0 &  0 \\
 1 & -1 & 1 & -1 & 1 &  0 &  0 &  0 \\
 0 &  0 & 1 & -1 & 1 &  0 &  0 &  0 \\
 0 &  0 & 1 & -1 & 1 & -1 &  1 &  0 \\
 0 &  0 & 0 &  0 & 1 & -1 &  1 &  0 \\
 0 &  0 & 0 &  0 & 1 & -1 &  1+g_r & -1-g_r \\
 0 &  0 & 0 &  0 & 0 &  0 &  g_r &  -g_r 
\end{array}
\!\right]
\eeas

\subsubsection{Boundary Conditions as Perturbations}

To study the effect of boundary conditions on the state transition
matrices $\AW$ and $\AK$, it is convenient to write the terminated
transition matrix as the sum of of the ``left-clamped'' case $\AWtt$
(for which $g_l=-1$) plus a series of one or more rank-one
perturbations.  For example, introducing a right termination with
reflectance $g_r$ can be written
\beqn
\AWttt = \AWtt + g_r\db_{8,7} = \AW - \db_{1,2} + g_r\db_{8,7},
\elabel{AWp}
\eeqn
where $\db_{ij}$ is the $M\times M$ matrix containing a 1 in its
$(i,j)$th entry, and zero elsewhere. (Following established
convention, rows and columns in matrices are numbered from 1.)  

In general, when $i+j$ is odd, adding $\db_{ij}$ to $\AWtt$
corresponds to a \emph{connection} from left-going waves to
right-going waves, or vice versa (see \fref{wglossless}).  When $i$ is
odd and $j$ is even, the connection flows from the right-going to the
left-going signal path, thus providing a termination (or partial
termination) on the right.  Left terminations flow from the bottom to
the top rail in \fref{wglossless}, and in such connections $i$ is even
and $j$ is odd.  The spatial sample numbers involved in the connection
are $2\lfloor (i-1)/2\rfloor$ and $2\lfloor (j-1)/2\rfloor$, where
$\lfloor x\rfloor$ denotes the greatest integer less than or equal to
$x$.

The rank-one perturbation of the DW transition matrix \eref{AWp}
corresponds to the following rank-one perturbation of the FDTD
transition matrix $\AKtt$:
\[
\AKttt \;\isdef\; \AKtt + g\Db_{8,7}
\]
where
\beqa
\Db_{8,7} &\isdef& \T\db_{8,7}\Ti
=
\left[\!
\begin{array}{rrrrrrrrrrr}
 0 & 0 & 0 & 0 &  0 & 0 & 0 & 0 \\
 \vdots & \vdots & \vdots & \vdots &  
 \vdots & \vdots & \vdots & \vdots\\
 0 & 0 & 0 & 0 &  0 & 0 & 0 & 0 \\
 0 & 0 & 0 & 0 &  0 & 0 & 1 & -1 \\
 0 & 0 & 0 & 0 &  0 & 0 & 1 & -1 
\end{array}
\!\right].
\elabel{Ddef}
\eeqa
In general, we have
\beqn
\Db_{ij} 
= \sum_{\kappa=j}^M
(-1)^{\kappa-j} \left(\db_{i\kappa}+\db_{i-1,\kappa}\right).
\elabel{grule}
\eeqn
Thus, the general rule is that $\db_{ij}$ transforms to a matrix
$\Db_{ij}$ which is zero in all but two rows (or all but one row when
$i=1$).  The nonzero rows are numbered $i$ and $i-1$ (or just $i$ when
$i=1$), and they are identical, being zero in columns $1:j-1$, and
containing $[1,-1,1,-1,\ldots]$ starting in column $j$.

\subsubsection{Reactive Terminations}

In typical string models for virtual musical instruments, the ``nut
end'' of the string is rigidly clamped while the ``bridge end'' is
terminated in a \emph{passive reflectance} $S(z)$.  The condition
for passivity of the reflectance is simply that its gain be bounded
by 1 at all frequencies \cite{SmithDWMMI}:
\beqn
\abs{S(\ejoT)}\leq 1, \quad \forall\, \omega\Ts\in[-\pi,\pi).
\elabel{sc}
\eeqn
A very simple case, used, for example, in the Karplus-Strong
plucked-string algorithm, is the two-point-average filter:
\[
S(z) = -\frac{1+\zi}{2}
\]
To impose this lowpass-filtered reflectance on the right in the chosen
subgrid, we may form
\[
\AWttt = \AWtt - \frac{1}{2}\Db_{8,5} - \frac{1}{2}\Db_{8,7}
\]
which results in the FDTD transition matrix
\beas
\AKttt &\isdef&
\left[\!
\begin{array}{ccccccccccc}
 0 & -1 & 1 & 0 &  0 &  0 &  0 &  0 \\
 1 & -1 & 1 &  0 & 0 &  0 &  0 &  0 \\
 1 & -1 & 1 & -1 & 1 &  0 &  0 &  0 \\
 0 &  0 & 1 & -1 & 1 &  0 &  0 &  0 \\
 0 &  0 & 1 & -1 & 1 & -1 &  1 &  0 \\
 0 &  0 & 0 &  0 & 1 & -1 &  1 &  0 \\
 0 &  0 & 0 &  0 & 1/2 & -1/2 &  0 &  0 \\
 0 &  0 & 0 &  0 & -1/2 &  1/2 & -1 & -1
\end{array}
\!\right].
\eeas
This gives the desired filter in a half-rate, staggered grid case.
In the full-rate case, the termination filter is really
\[
S(z) = -\frac{1+\zmt}{2}
\]
which is still passive, since it obeys \eref{sc}, but it does not have
the desired amplitude response: Instead, it has a notch (gain of 0)
at one-fourth the sampling rate, and the gain comes back up to 1 at
half the sampling rate.  In a full-rate scheme, the two-point-average
filter must straddle both subgrids.

Another often-used string termination filter in digital waveguide
models is specified by \cite{SmithDWMMI}
\beas
s(n) &=& -g\left[\frac{h}{4}, \frac{1}{2}, \frac{h}{4}\right]\\
\longleftrightarrow\quad S(\ejoT)&=& 
-e^{-j\omega \Ts}g\frac{1 + h \cos(\omega \Ts)}{2},
\eeas
where $g\in(0,1)$ is an overall gain factor that affects the decay
rate of all frequencies equally, while $h\in(0,1)$ controls the
relative decay rate of low-frequencies and high frequencies.  An
advantage of this termination filter is that the delay is
always one sample, for all frequencies and for all parameter settings;
as a result, the tuning of the string is invariant with respect to
termination filtering.  In this case, the perturbation is
\[
\AWttt = \AWtt - \frac{gh}{4}\delta(M-5,M) 
	      -  \frac{g}{2}\delta(M-3,M)
              - \frac{gh}{4}\delta(M-1,M) 
\]
and, using \eref{grule}, 
the order $M=8$ FDTD state transition matrix is given by
\beas
\AKttt &\isdef&
\left[\!
\begin{array}{ccccccccccc}
 0 & -1 & 1 & 0 &  0 &  0 &  0 &  0 \\
 1 & -1 & 1 &  0 & 0 &  0 &  0 &  0 \\
 1 & -1 & 1 & -1 & 1 &  0 &  0 &  0 \\
 0 &  0 & 1 & -1 & 1 &  0 &  0 &  0 \\
 0 &  0 & 1 & -1 & 1 & -1 &  1 &  0 \\
 0 &  0 & 0 &  0 & 1 & -1 &  1 &  0 \\
 0 &  0 & g_1 &  -g_1 & 1+g_2 & -1-g_2 & 1+g_3 & -1-g_3 \\
 0 &  0 & g_1 &  -g_1 & \quad g_2 & \quad -g_2 & \quad  g_3 & \quad  -g_3
\end{array}
\!\right]
\eeas
where
\beas
g_1 &\isdef& -\frac{gh}{4}\\
g_2 &\isdef& -\frac{g}{2}+g_1\\
g_3 &\isdef& -\frac{gh}{4}+g_2.\\
\eeas


The filtered termination examples of this section generalize
immediately to arbitrary finite-impulse response (FIR) termination
filters $S(z)$.  Denote the impulse response of the termination filter
by
\[
s(n)=[s_0,s_1,s_2,\ldots,s_N],
\]
where the length $N$ of the filter does not exceed $M/2$.  Due to
the DW-FDTD equivalence, the general stability condition is stated
very simply as
\[
\abs{S(\ejoT)} = \abs{\sum_{n=0}^{N-1} s_n e^{-j\omega \Ts}} \leq 1, 
\quad \forall\, \omega\Ts\in[-\pi,\pi).
\]

\subsubsection{Interior Scattering Junctions}

A so-called \emph{Kelly-Lochbaum scattering junction}
\cite{MG,SmithDWMMI} can be introduced into the string at the fourth
sample by the following perturbation
\[
\AKtttt = \AKtt + 
(1-k_l)\Db_{5,3} + 
   k_r \Db_{5,8} + 
   k_l \Db_{6,3} + 
(1-k_r)\Db_{6,8}.
\]
Here, $k_l$ denotes the reflection coefficient ``seen'' from left to
right, and $k_r$ is the reflectance of the junction from the right.
When the scattering junction is caused by a change in string density
or tension, we have $k_r=-k_l$.  When it is caused by an externally
imposed termination (such as a plectrum or piano-hammer touching the
string), we have $k_r=k_l$, and the reflectances may become filters
instead of real values in $[-1,1]$.  Energy conservation demands that
the transmission coefficients be amplitude complementary with respect
to the reflection coefficients \cite{SmithDWMMI}.


A single time-varying scattering junction provides a reasonable model
for plucking, striking, or bowing a string at a point.  Several
adjacent scattering junctions can model a distributed interaction,
such as a piano hammer, finger, or finite-width bow spanning several
string samples.

Note that scattering junctions separated by one spatial sample (as
typical in ``digital waveguide filters'' \cite{SmithDWMMI}) will
couple the formerly independent subgrids.  If scattering junctions are
confined to one subgrid, they are separated by two samples of delay
instead of one, resulting in round-trip transfer functions of the form
$H(z^2)$ (as occurs in the digital waveguide mesh).  In the context of
a half-rate staggered-grid scheme, they can provide general IIR
filtering in the form of a ladder digital filter \cite{MG,SmithDWMMI}.

\subsection{Lossy Vibration}

The DW and FDTD state-space models are equivalent with respect to
lossy traveling-wave simulation. \Fref{wglossy} shows the flow diagram
for the case of simple attenuation by $g$ per sample of wave
propagation, where $g\in(0,1]$ for a passive string.

\myTexFigure{wglossy}{DW flow diagram in the lossy case.}

The DW state update can be written in this case as
\[
\xW(n+2) = g^2\AW\xW(n) + \BW \uv(n+2).
\]
where the loss associated with two time steps has been incorporated
into the chosen subgrid for physical accuracy.  (The neglected subgrid
may now be considered lossless.)  In changing coordinates to the FDTD
scheme, the gain factor $g^2$ can remain factored out, yielding
\[
\xK(n+2) = g^2\AK\xK(n) + \BK \uv(n+2).
\]
When the input is zero after a particular time, such as in a plucked
or struck string simulation, the losses can be implemented at the
final output, and only when an output is required, \eg,
\[
y(n) = g^n y_0(n)
\]
where $y_0(n)$ denotes the corresponding lossless simulation.  When
there is a general input signal, the state vector needs to be properly
attenuated by losses.  In the DW case, the losses can be lumped at two
points per spatial input and output \cite{SmithDWMMI}.

\subsection{State Space Summary}

We have seen that the DW and FDTD schemes correspond to state-space
models which are related to each other by a simple change of
coordinates (similarity transformation).  It is well known that such
systems exhibit the same transfer functions, have the same modes, and
so on.  In short, they are the same linear dynamic system.
Differences may exist with respect to spatial locality of input
signals, initial conditions, and boundary conditions.

State-space analysis was used to translate initial conditions and
boundary conditions from one case to the other.  Passive terminations
in the DW paradigm were translated to passive terminations for the
FDTD scheme, and FDTD excitations were translated to the DW case in
order to interpret them physically.

\section{Computational Complexity}

The DW model is more efficient in one dimension because it can make
use of delay lines to obtain an $\Oscr(1)$ computation per time sample
\cite{SmithPMUDW}, whereas the FDTD scheme is $\Oscr(M)$ per sample
($M$ being the number of spatial samples along the string).  There is
apparently no known way to achieve $\Oscr(1)$ complexity for the FDTD
scheme.  In higher dimensions, \ie, when simulating membranes and
volumes, the delay-line advantage disappears, and the FDTD scheme has
the lower operation count (and memory storage requirements).

\section{Summary}

An explicit linear transformation was derived for converting state
variables of the finite-difference time-domain (FDTD) scheme to those
of the digital waveguide (DW) scheme.  The equivalence of the FDTD and
DW state transitions was reviewed, and the proof of state-space
equivalence was completed.  Since the DW scheme is exact within its
bandwidth (being a sampled traveling-wave scheme instead of a finite
difference scheme), it can be put forth as the proper physical
interpretation of the FDTD scheme, and consequently be used to provide
physically accurate initial conditions and excitations for the FDTD
method.  For its part, the FDTD method provides lower cost relative to
the DW method in dimensions higher than one (for simulating membranes,
volumes, and so on), and can be preferred in highly distributed
nonlinear string simulation applications.

\section{Future Work}

The simple state translation formulas derived here for the
one-dimensional case do not extend simply to higher dimensions.  While
straightforward extensions to higher dimensions are presumed to exist,
a simple and intuitive result such as found here for the 1D case could
be more useful for initializing and driving FDTD mesh simulations from
a physical point of view.  In particular, spatially localized initial
conditions and boundary conditions in the DW framework should map to
localized counterparts in the FDTD scheme.  A generalization of the
Toeplitz operator $\T$ having a known closed-form inverse could be
useful in higher dimensions.

{\small
\raggedright

}

\end{document}